\documentclass{ws-procs961x669}            
\usepackage{psfrag}
\usepackage{graphicx}

\newcommand{\pd}{\partial}
\newcommand\m{\mu}

\newcommand\n{\nu}

\newcommand\s{\sigma}
\renewcommand\a{\alpha}
\renewcommand\b{\beta}

\def\be{\begin{equation}}
\def\ee{\end{equation}}
\def\bea{\begin{eqnarray}}
\def\eea{\end{eqnarray}}

\begin{document}

\title{Modified gravity
and binary pulsars: the Lorentz violating case}

\author{D. Blas$^*$}

\address{Theoretical Physics Department, CERN,  Geneva, Switzerland.\\
$^*$E-mail: diego.blas@cern.ch}

\begin{abstract}
\vspace{-4.5cm}
\begin{flushright}
CERN-TH-2016-014
\end{flushright}
\vspace{3.5cm}
\noindent
The dynamics of binary pulsars can be used to test different aspects of gravitation. 
This is particularly important to constrain alternatives to general relativity in regimes which are not
probed by other methods. In this
short contribution, I will describe the case of theories of gravity without Lorentz invariance. 
The latter are important in the context of  quantum gravity 
and modify the laws of gravity at basically all scales.
\end{abstract}


\bodymatter

\section{Introduction}\label{aba:sec1}

General Relativity (GR) is living a golden era of continuous verifications of its predictions, at many scales and in very different processes \cite{Will:2014bqa,Berti:2015itd,Clifton:2011jh}. The agreement of data with 
GR predictions is both astonishing (given the range of scales probed) and disappointing (GR is not a complete quantum theory, 
but the experimental guidance towards its completion is scarce). 

One hopes that the synergy between the experimental and theoretical efforts will unveil the remaining 
mysteries of gravitation. In this note I shall illustrate this  by working with a concrete model 
motivated by  quantum gravity, Lorentz violating (LV) gravity, and that modifies the predictions of GR
at basically all scales. I will describe the model and focus on the predictions for binary pulsars.

For more information, the reader is invited to consult the 
recent review article \cite{Blas:2014aca}.

\section{Lorentz violating gravity as a theory of modified gravity}

The number of theories of gravity beyond GR has experienced a remarkable recent expansion.
 This was driven by the 
new data, the continuous search for theories that may solve the shortcomings of GR and the advances in the methods to work in problems related to gravitation. The attempt to classify the different approaches has produced a relatively complex map of possibilities \cite{Will:2014bqa,Berti:2015itd,Clifton:2011jh}. Instead of classifying them, one can simply confront them with the following list of desirable properties for a theory modifying GR: 
\begin{romanlist}[(iii)]
\item Represent/break a fundamental principle of gravity/Nature.
\item Improve the short distance properties of GR (ideally addressing the problems of quantum gravity and/or singularities).
\item Provide new ideas for cosmic acceleration or dark matter.
\item Produce a (testable) interesting phenomenology.
\end{romanlist}
We are interested in theories where 
Lorentz invariance is broken by a preferred local time direction. The latter is
represented by a time-like vector field $u^\mu$ satisfying
\be
\label{eq:unit}
u_\mu u^\mu=1.
\ee
Point (i) is clearly met: Lorentz invariance is one of the fundamental assumptions
of the modern theories of particle physics and gravitation. To study the evidence for such hypothesis, 
one needs to understand the experimental constraints on possible deviations \cite{Liberati:2013xla}. 
Point (ii) is of particular importance and LV gravity has a special status in this regard. First, 
even when Lorentz invariance is considered a fundamental symmetry, 
 the complete theory of (quantum) gravity may admit low-energy phases 
where it is broken spontaneously. This approach is known as 
 Einstein-aether theory \cite{Jacobson:2000xp}. Furthermore, if the vector
$u^\m$ corresponds to the normal of a foliation of space-time into space-like hypersurfaces 
\be
\label{eq:kh}
u_\m\equiv\frac{\pd_\mu \varphi}{\sqrt{g^{\m\n}\pd_\m  \varphi\pd_\n \varphi}},
\ee
the theory may be embedded in a renormalizable theory known as Ho\v rava gravity \cite{Horava:2009uw,Blas:2009qj,Barvinsky:2015kil}.  The consequences for black hole singularities are not completely understood (see however \cite{Blas:2014aca}). The generic theories defined with a vector of the form (\ref{eq:kh}) are known as khronometric theories. Regarding point (iii), the presence of the field $u^\mu$  allows for new natural potentials that can drive
the current cosmic acceleration \cite{Blas:2011en} and present new ideas for models of modified Newtonian dynamics \cite{Bonetti:2015oda}. Finally, since Lorentz invariance is a gauge symmetry in GR, new massless excitations can be excited when it is broken. Those change the predictions of GR at all scales \cite{Blas:2014aca}.

To construct the action of LV theories of gravity, one writes the different operators including $u^\mu$ and $g^{\m\n}$, covariant under diffeomorphisms and organized in a derivative expansion  (we also assume CPT),
\be 
\label{ae-action} 
S = -\frac{M_0^2}{2}\int  ~d^{4}x \sqrt{-g}~ \left(R
+K^{\a\b}{}_{\m\n} \nabla_\a u^\m \nabla_\b u^\n+\lambda (u^\m u_\m-1)+\frac{{\mathcal O^{n+2}}}{M^n_\star}\right)\,,
\ee 
where $g$ and $R$ are the metric determinant and the Ricci scalar and 
\be
\label{eq:K}
K^{\a\b}{}_{\m\n} \equiv c_1 g^{\a\b}g_{\m\n}+c_2\delta^{\a}_{\m}\delta^{\b}_{\n}
+c_3 \delta^{\a}_{\n}\delta^{\b}_{\m}+c_4 u^\a u^\b g_{\m\n}\, .
\ee 
We use the constant $M_0$ instead of $M_P$ for the mass scale
in front of the Einstein-Hilbert action to distinguish it from the quantity appearing in Newton's law \eqref{eq:Newton}. 
By the last term in \eqref{ae-action}  we indicate the higher dimensional operators, which we assume to be suppressed by a scale $M_\star$. These operators are negligible for
the astrophysical observations we will consider \cite{Yagi:2013ava} and we will work in the limit $M_\star\to \infty$.
We imposed the restriction (\ref{eq:unit}) through a Lagrange multiplier $\lambda$. In the khronometric
case this is not necessary. If the condition \eqref{eq:kh} is satisfied,  one of the terms in (\ref{eq:K})  can be expressed in terms
of the others. In particular, one can absorb the $c_1$ term into the second, third and forth term with respective coupling  constants
\be\label{eq:EAtoKH}
\lambda\equiv c_2, \quad \beta\equiv c_3+c_1, \quad \alpha\equiv c_4+c_1.
\ee
The action \eqref{ae-action} can be reformulated in terms of geometrical quantities of the congruences of  $u^\mu$, see \cite{Jacobson:2013xta}. The operators in this action are multiplied by the parameters
\be
\label{eq:cdef}
c_\theta\equiv c_1+c_3+3c_2, \quad c_\sigma\equiv c_1+c_3, \quad c_\omega\equiv c_1-c_3,\quad c_a\equiv c_1+c_4,
\ee
which we will use henceforth. This is convenient because the predictions of the khronometric theory can be found from the Einstein-aether calculations in the limit 
\be
\label{eq:untwist}
c_\omega\to \infty.
\ee

\section{Degrees of freedom and matter fields}\label{sec:dofs}

The configuration  $g_{\m\n}=\eta_{\m\n}$ and $u^\m=\delta^\m_0$ is a solution of the equations of motion derived from  the action \eqref{ae-action}. The spectrum of perturbations around it consists of a tensor graviton mode, a vector and a scalar modes characterised by the velocities
\be
\label{eq:speeds}
v_t^2= \frac{1}{1-c_\sigma}, \quad v_v^2=\frac{c_\sigma+c_\omega-c_\sigma c_\omega}{2c_\sigma(1-c_\sigma)}, \quad v_s^2=\frac{(c_\theta+2c_\sigma)(1-c_a/2)}{3c_a(1-c_\sigma)(1+c_\theta/2)}.
\ee
Notice that the graviton's velocity is modified. These velocities should all be real to avoid gradient instabilities. The absence of gravitational Cherenkov radiation requires that they are close to or greater than the speed of light. In the khronometric limit \eqref{eq:untwist}, the vector modes disappear (they become infinitely heavy).

Before discussing the phenomenological constraints in the previous parameters, it is worth mentioning that the most serious challenges of LV theories come from the matter sector. In fact, the field $u_\mu$ can couple to the fields in the standard model of particle physics (SM). This generates LV effects whose study has 
produced very strong bounds on the possible couplings \cite{Kostelecky:2008ts,Liberati:2013xla}.
The consequences of these bounds for gravitation are not universal.  If  LV affects identically the SM and gravitation the smallness of the parameters $c_i$ will be beyond any foreseeable test. However, there are mechanisms that separate the LV effects in these sectors and make the SM tests irrelevant to constrain $c_i$.  This  may happen  due to (softly broken) supersymmetry, 
 suppression mechanisms 
 or by a dynamical emergence of LI at low energies in the SM  (see \cite{Liberati:2013xla} for review).
We will assume that the fields in the SM do not interact directly with the field $u_\m$.

\section{Solar System constraints}\label{sec:solar}

The LV parameters appearing in \eqref{ae-action}  are efficiently constrained by
Solar System observations. For the latter, it is enough to consider the post-Newtonian (PN) limit of the
equations of motion  \cite{Will:2014xja}. Unlike in GR, one needs to specify the velocity of the sources (the Sun in this case) with respect to the preferred frame represented by $u^\mu$ (in particular its PN order). The most common assumption is that the preferred frame is basically aligned with the cosmic microwave background (CMB) frame. This  is in agreement with requiring isotropy of the Universe at large scales. 
Under these conditions the Newton's constant appearing in Newton's law  reads
\be
\label{eq:Newton}
G_N\equiv \frac{1}{4\pi M_0^2 \left(2-c_a\right)},
\ee
and the only PPN parameters which differ from GR are
\be
\begin{split}
\label{eq:PPN}
&\alpha_1=4\left(\frac{c_\omega(c_a-2c_\sigma)+c_a c_\s}{c_\omega(c_\s
-1)-c_\s}\right)\,, \quad \alpha_2= \frac{\alpha_1}{2} +\frac{3(c_a-2 c_\sigma)(c_\theta+c_a)}{(2-c_a)(c_\theta+2c_\sigma)}\,.
\end{split}
\ee
One can take the limit \eqref{eq:untwist} for the khronometric case.
The  tests of gravitation coming from lunar laser ranging
and solar alignment with the ecliptic ~\cite{Will:2014xja,Will:2014bqa} yield constraints
for the deviations of GR of the form $|\alpha_1|\lesssim 10^{-4}$ and
$|\alpha_2| \lesssim 10^{-7}$. These conditions only constraint two combinations of the $c_i$ coefficients. 

\section{Strongly gravitating objects }\label{sec:strong}

\subsection{Conservative dynamics}

Lorentz violating effects are important for the dynamics of compact objects.  
The coupling between the field
$u^\m$ and the gravitational fields is  relevant only close to the object (where the gravitational fields are stronger).  For the physics far away from the source, this coupling can be thus described as an
effective interaction of the whole body with the aether field  (see
\cite{Damour:1992we} for similar effects in other   theories). In this approximation the action of the source can be described as a point-particle with an aether charge 
\be
\label{eq:acsens}
S_{\rm pp \; A}=-\int ds_A \tilde m_A(\gamma_A), 
\ee
where $\tilde m_A$ is a function of the Lorentz factor of the source with respect to the aether, $\gamma_A\equiv u_\m v_A^\m$, with $v_A^\m$ being the four-velocity of the source. Finally, $ds_A$ is the line element of the trajectory. Assuming  that $\gamma_A$ is close to one, 
\begin{align}
\label{taylor-pp}
S_{\rm pp \; A} &=- \tilde{m}_{A}  \int ds_A \; 
\left\{1 + \sigma_{A} \left(1- \gamma_{A}\right)
+ {\cal{O}}\left[\left(1-\gamma_{A}\right)^{2}\right] \right\}\,, 
\end{align}
where we have defined the sensitivity parameters $\sigma_{A}$ via %
\begin{align}
\label{sigma-def}
\sigma_{A} &\equiv - \left.\frac{d \ln \tilde{m}_{A}(\gamma_A)}{d \ln \gamma_{A}}
\right|_{\gamma_{A} = 1}\,.
\end{align}
The previous action  can be used to describe the dynamics of  systems with isolated objects. 
The coupling between the stars and $u^\m$ 
violates  the equivalence principle and implies that the momentum defined in GR is not
conserved in these theories.

Let us consider  the gravitational field  of an isolated object far away from it.
  At the PN level, the metric will be of the PPN form, but this time, and in contrast to GR, the PPN parameters will depend on the internal structure of the body through the sensitivities 
\cite{Damour:1992we,Will:2014xja}. This follows from the violation of the \emph{strong} equivalence principle.  
The {\it strong} PPN parameters  $\hat \alpha_i$ read  \cite{Foster:2007gr,Yagi:2013ava}
\begin{align}
\label{SF-alpha1}
\hat{\alpha}_{1} &= \alpha_1 + \frac{c_\omega (8+\alpha_1) \sigma_A}{2 c_1}\, , \quad
\hat{\alpha}_{2} = \alpha_2 + \hat \alpha_1-\alpha_1- \frac{  (c_a -2)  (\alpha_1 - 2 \alpha_2) \sigma_A}{2  (c_a - 2 c_\sigma)}\,.
\end{align} 
 These quantities can be constrained very efficiently with data from binary and isolated pulsars \cite{Weisberg:2013pha,Shao:2013wga}. By using the orbital dynamics of the pulsar-white
dwarf binary PSR J1738+0333, one  gets $\left|\hat \alpha_1\right|< 10^{-5}\quad (95\%\ \mathrm{CL})$ . Using data from solitary pulsars one can get the bounds 
$\left|\hat \alpha_2\right|< 1.6\cdot 10^{-9}\quad (95\%\ \mathrm{CL})$. 

These bounds are very strong, but they are source dependent and to translate them into constraints for the fundamental parameters of \eqref{ae-action} one needs the sensitivities $\sigma_A$.
This requires solving the field equations for realistic sources and matching the metric to the PPN form in the region where the PN expansion applies. 
The analytical calculation in the weak-field limit appeared in \cite{Foster:2007gr}. 
This approximation is  not   applicable in situations with 
very compact objects. 
In Refs.  \cite{Yagi:2013qpa,Yagi:2013ava}, the metric outside  spherically-symmetric, non-rotating, cold (and thus old) neutron stars was computed for different equations of state. It was also assumed $|\gamma_A-1|\ll 1$.  The objects for which the previous approximation holds are well represented  within the current data set of pulsars (recall that 
$u^\mu$ is almost aligned with the CMB frame). At zeroth order in this parameter the profile of the star is modified (different Tolman-Oppenheimer-Volkoff equations).  These modifications are not enough to produce strong constraints.
 By comparing the numerical solution at first order in $|\gamma_A-1|$ with the asymptotic PPN metric one can find  $\sigma_A$ and  $\hat \alpha_i$ \cite{Yagi:2013qpa,Yagi:2013ava}.
  To extract the sensitivities for real objects one needs to  know their mass, which can be found  in binary systems, but not for isolated pulsars.
The dynamics of the  pulsar-white
dwarf binaries PSR J1738+0333 \cite{Freire:2012mg} already put very strong constraints in the LV parameters, breaking the degeneracies remaining after imposing the
 Solar System bounds \cite{Yagi:2013ava}. These constraints are included in Fig.~\ref{fig3}.

\subsection{Dissipative effects: emission of gravitational waves}\label{sec:emission}
 
Another  key  test  of GR comes from the damping of orbits of binary systems due to the emission of gravitational waves (GWs) \cite{Will:2014xja}.  The modifications  in  LV theories come from 
the different dynamics and the changes in the properties (and number) of the radiating degrees for freedom. 
 Their study was started in \cite{Foster:2007gr} and was completed in \cite{Yagi:2013qpa,Yagi:2013ava}. The first important observation  is that the change in the orbital period of the system contains a dipolar term which is enhanced with respect to the GR quadrupolar term by ${\cal O}(c/v)^2$, where $v$ is the typical orbital velocity. This is related to the violation of the strong equivalence principle. This contribution is multiplied by the difference of the sensitivities squared, which means that it will be important only for asymmetric systems (e.g. a binary of a neutron star and a white dwarf).  
The quadrupolar contribution is modified by both the speed of the graviton being $v_2$ and the presence of other degrees of freedom.
\begin{figure}
\begin{center}
\includegraphics[width=.8\textwidth]{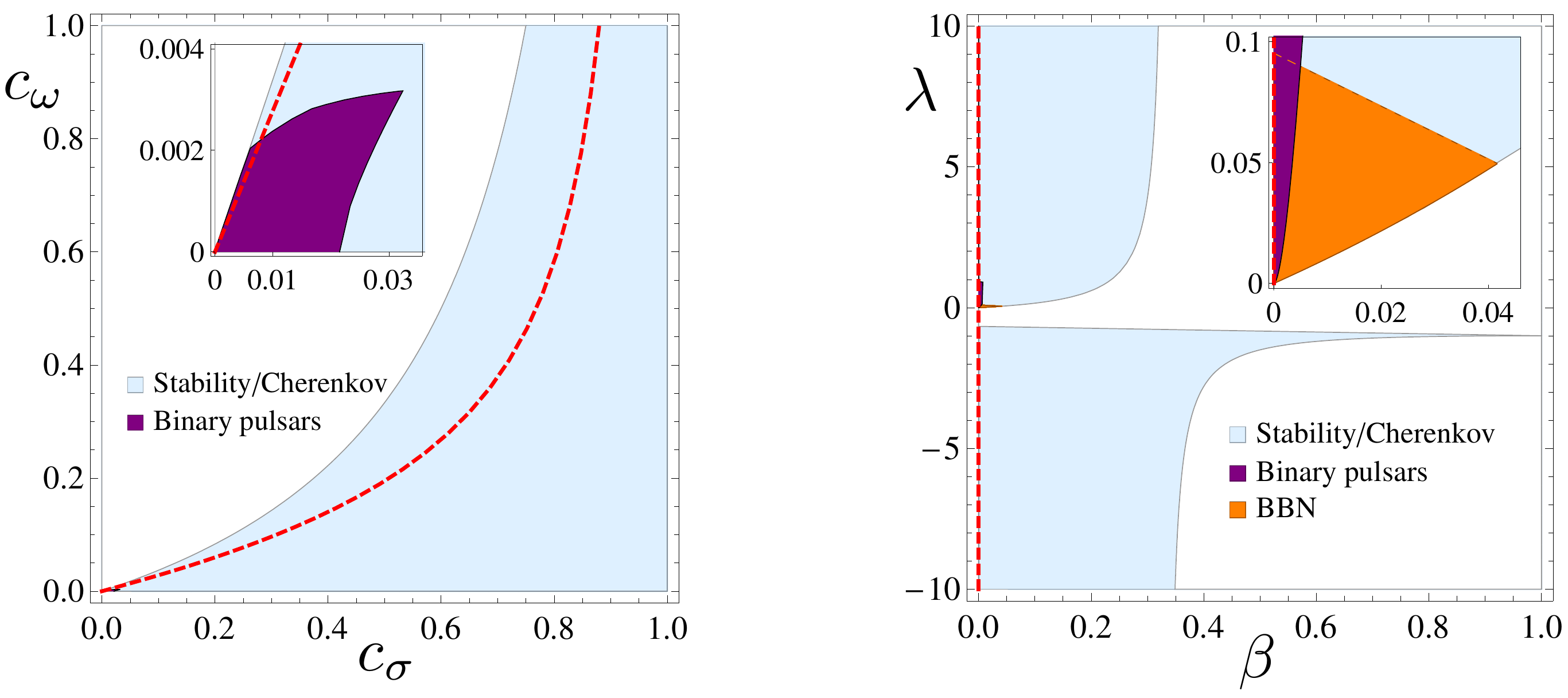}
\caption{\label{fig3}
Constraints from pulsars on LV parameters unconstrained by Solar system tests. Left:  Einstein-aether; Right: khronometric. The areas outside shaded regions are ruled out by stability/Cherenkov (light blue), BBN (orange) and pulsar constraints (purple). Red-dotted line: values for which the orbital decay rate from GWs agrees with GR in the weak limit. 
Figures from \cite{Yagi:2013ava}.
}
\end{center}
\end{figure}
By comparing the predictions for LV theories to the systems   PSR J1141-6545, PSR J0348+043 and PSR J0737-3039, adding the constraints from the \emph{strong} PPN parameters and Solar System tests one gets the constraints shown in Fig.~\ref{fig3}.
 Besides being very stringent, and contrary to Solar System tests,  the observations from pulsars constrain \emph{all} the $c_i$ parameters. Once these constraints are imposed,  LV theories are completely viable.

\section{Conclusion}\label{sec:con}

Theories of gravity that violate Lorentz invariance are well motivated alternatives to GR with better
properties at high energies. In this short contribution I have introduced them and discussed some phenomenological implications. In
particular, I have shown how binary pulsars observations are very useful to constrain these theories in regimes not explored
by other observations and to break degeneracies of other tests in the parameter space. These constraints are summarized in Fig.~\ref{fig3}.


\vspace{-.1cm}




\end{document}